\documentclass[%
 reprint,
 superscriptaddress, showkeys,
 amsmath,amssymb,
 aps, physrev,
]{revtex4-2}
\usepackage{siunitx}
\usepackage{xcolor}
\usepackage{graphicx}
\usepackage{dcolumn}
\usepackage{bm}

\begin{document}

\title{Enhanced and directional light emission from two-dimensional excitons using Mie voids}

\author{Avishek Sarbajna}
\affiliation{Department of Physics, Technical University of Denmark, Fysikvej, Kongens Lyngby, DK-2800 Denmark}

\author{Ganesh Ghimire}
\affiliation{Department of Physics, Technical University of Denmark, Fysikvej, Kongens Lyngby, DK-2800 Denmark}

\author{Ilia D. Breev}
\affiliation{Department of Physics, Technical University of Denmark, Fysikvej, Kongens Lyngby, DK-2800 Denmark}

\author{Xavier Zambrana-Puyalto}
\affiliation{Department of Physics, Technical University of Denmark, Fysikvej, Kongens Lyngby, DK-2800 Denmark}

\author{Cheng Xiang}
\affiliation{Department of Physics, Technical University of Denmark, Fysikvej, Kongens Lyngby, DK-2800 Denmark}

\author{Alexander Huck}
\affiliation{Department of Physics, Technical University of Denmark, Fysikvej, Kongens Lyngby, DK-2800 Denmark}

\author{Timothy J. Booth}
\affiliation{Department of Physics, Technical University of Denmark, Fysikvej, Kongens Lyngby, DK-2800 Denmark}

\author{S{\o}ren Raza}
\email[Corresponding author S{\o}ren Raza, email: ]{sraz@dtu.dk}
\affiliation{Department of Physics, Technical University of Denmark, Fysikvej, Kongens Lyngby, DK-2800 Denmark}



\begin{abstract}
Controlling light emission at the nanoscale has important applications in solid-state lighting, displays, and quantum light sources. Achieving this control requires both enhanced local electromagnetic fields to boost emission intensity and engineered radiation patterns to direct photons efficiently. Mie voids, consisting of an air cavity surrounded by a high-index semiconductor, are particularly suited for this purpose because they expose their strongest fields in an accessible region for nearby emitters while supporting resonances that shape directional emission through interference. Here, we demonstrate an all–van der Waals nanophotonic platform that couples excitons in atomically thin WS$_2$ to Mie void resonators formed in WSe$_2$. Guided by electromagnetic simulations, we identify void geometries that maximize photoluminescence through synergistic enhancement of excitation and emission processes. We also develop a two-step fabrication strategy that enables independent control of void diameter and depth, providing a route to systematically tune the optical response. Experimentally, we observe up to a 600-fold increase in photoluminescence intensity from monolayer WS$_2$ placed on individual voids compared to on an unstructured WSe$_2$, along with pronounced out-of-plane beaming of light that yields a forward-to-off-axis enhancement of 2.6~dB. Our results establish Mie voids in van der Waals semiconductors as a new platform for controlling light–matter interactions and realizing compact, directional, and efficient nanoscale light sources.
\end{abstract}

\maketitle


\section*{Introduction}
Controlling the emission of light at the nanoscale is central to modern photonics, with applications ranging from solid-state lighting to single-photon quantum sources. The ability to enhance the emission intensity and control the direction of emission opens routes to highly efficient and compact light sources. Optical nanoantennas provide a versatile means to achieve this control, offering subwavelength confinement of electromagnetic fields and precise manipulation of emission characteristics~\cite{Novotny2011,Li2020}. This control relies on two underlying mechanisms, namely the amplification of the local electromagnetic field experienced by the emitter and the tailoring of the far-field radiation pattern so that photons are efficiently directed toward the desired collection direction.

Early realizations of optical nanoantennas relied on metallic structures that confine light below the diffraction limit through plasmonic resonances~\cite{Curto2010, Chaubey2021, Tong2013, Shegai2011, Palacios2017, Coenen2014}. While highly effective at boosting local fields, plasmonic antennas suffer from intrinsic absorption losses and often require complex geometries. To overcome these limitations, dielectric nanoantennas based on high-refractive-index materials have emerged as an attractive low-loss alternative~\cite{Kuznetsov2016,Staude2013,Fu2013, Person2013,Staude2013}. These resonators support both electric and magnetic Mie resonances that can enhance photoluminescence (PL)~\cite{Huang2022, Sortino2019, Shinomiya2022, Katrisioti2025, Poumirol2020} and interfere with the emitter’s dipole radiation to enable directional emission~\cite{Peter2017Directional, Cihan2018,Vaskin2018, Fang2021, Ozawa2025, Bidault2019, Murai2020, Bucher2019}. However, in conventional dielectric Mie resonators, the strongest electromagnetic fields are confined inside the high-index material, which makes it difficult to overlap the field maximum with nearby emitters~\cite{Lepeshov2019}. This limitation becomes even more pronounced because achieving optimal directionality typically requires either a significant emitter–resonator separation~\cite{Rolly2012} or the use of moderate-index materials~\cite{Rolly2013,Zhang2015}, both of which reduce the local field enhancement experienced by the emitter. Achieving simultaneous local-field amplification and directional emission therefore remains a central challenge.
\begin{figure*}[tp]
    \centering
    \includegraphics[width=\textwidth]{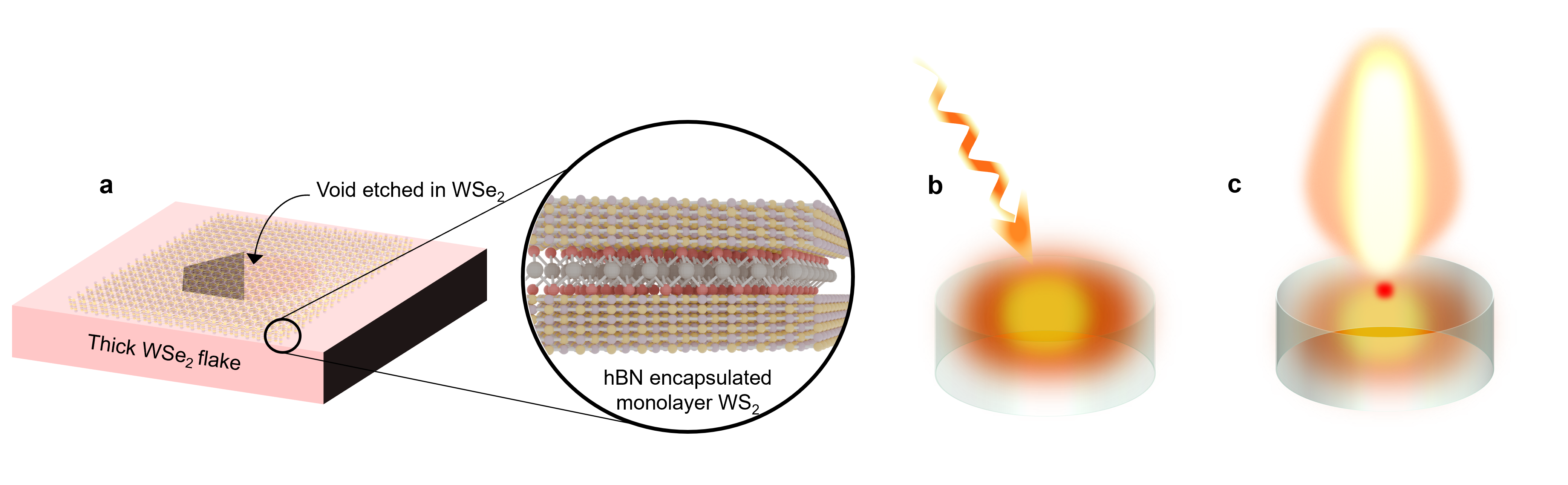}
    \caption{\textbf{Controlling light emission using void resonators.} \textbf{a,} Schematic of the device displaying an hBN encapsulated monolayer of WS$_2$ (zoomed) on top of a void resonator constructed in WSe$_2$. \textbf{b,} Field localization within the air region enabled by the void resonance. \textbf{c,} Emitter-void interaction that results in enhanced and out-of-plane directional emission.}
    \label{fig:1}
\end{figure*}

A promising route to address this challenge is to invert the conventional Mie resonator geometry and form structures known as Mie voids~\cite{Hentschel2023, Sarbajna2024}. These consist of a low-index region, typically air, surrounded by a high-index medium. Such complementary architectures support the same family of Mie resonances, namely electric and magnetic multipoles, but they expose their strongest electromagnetic fields inside the low-index region~\cite{Hamidi2025}. This geometry naturally places the electromagnetic field maximum in an accessible volume where an emitter can be positioned while maintaining the ability to tailor far-field emission through interference between the emitter and the void resonances. Since the optical confinement occurs within the void rather than the high-index host, Mie voids also allow the use of materials with high refractive index and moderate absorption, which further increases the index contrast and broadens the range of suitable materials. These combined properties make Mie voids an appealing platform for achieving simultaneous local-field enhancement and directional light emission.

In this work, we demonstrate the potential of Mie voids for controlling light emission using an all–van der Waals nanophotonic platform. We couple an atomically thin layer of tungsten disulfide (WS$_2$), encapsulated in few-nanometer-thick hexagonal boron nitride (hBN), to Mie void resonators formed in tungsten diselenide (WSe$_2$) (see Fig.~\ref{fig:1}a). Electromagnetic simulations show that carefully tuning the void dimensions leads to pronounced enhancement of PL through two synergistic effects, namely strong field confinement inside the void (Fig.~\ref{fig:1}b) and enhanced out-of-plane emission enabled by emitter–void coupling (Fig.~\ref{fig:1}c). We verify these predictions experimentally and observe up to a 600-fold enhancement in the PL intensity of the A-exciton in WS$_2$ coupled to individual voids at room temperature, compared to emission from unpatterned WSe$_2$. To demonstrate directional emission, we measure the angular distribution of emitted light using back-focal-plane imaging of void arrays and find that the emission profile evolves from the Lambertian pattern of uncoupled emitters to a highly directional distribution with enhanced emission normal to the surface. The combination of field localization in air, directional emission, and emitter–mode alignment provides a promising route toward compact and efficient light sources based on Mie voids in van der Waals materials.

\section*{Results}

\subsection*{Designing Mie voids for maximal light emission}

\begin{figure*}[tp]
    \centering
    \includegraphics[width=\textwidth]{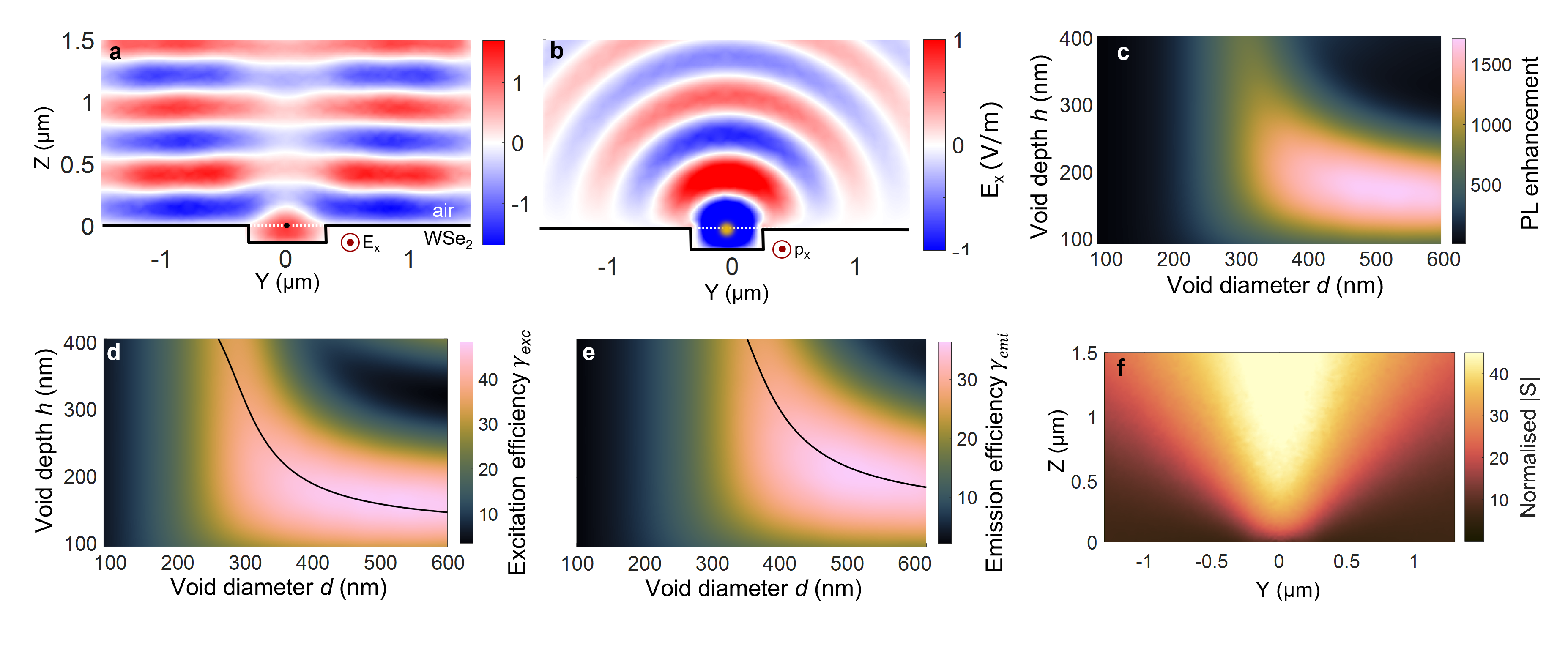}
    \caption{\textbf{Simulation of enhanced and directional light emission.} \textbf{a,} Simulated electric field distribution \(E_x\) in the $yz$-plane for a void with diameter $d=540$~nm and depth $h=160$~nm. The excitation is a normally-incident plane wave polarized along $x$ with a wavelength of $\lambda_\mathrm{exc}=532$~nm. The black border separates the air and WSe$_2$ domains. The white dashed line marks the top surface of the void. The black dot denotes the center of the void opening, where the electric field intensity is evaluated to obtain the excitation efficiency. \textbf{b,} Simulated electric field distribution \(E_x\) from an $x$-oriented dipole (yellow dot) emitting at a wavelength of $\lambda_\mathrm{emi}=610$~nm and placed at the center of the void opening. The void dimensions are $d=500$~nm and $h=200$~nm. \textbf{c,} Photoluminescence enhancement as a function of void diameter $d$ and depth $h$ calculated using Eq.~(\ref{eq:PLenh}). \textbf{d–e,} Excitation and emission efficiencies as a function of void dimensions, respectively. The black lines depict the depth vs. diameter relation given by Eq.~(\ref{eq:constinter}). \textbf{f,} Simulated Poynting vector amplitude of dipole-on-void system normalized by dipole on flat WSe$_2$. The void dimensions are the same as in \textbf{b}.}
    \label{fig:2}
\end{figure*}



The PL signal from monolayer WS$_2$ originates from an excitation process at a wavelength of $\lambda_\mathrm{exc} = 532$~nm, where incident photons generate electron–hole pairs in the conduction and valence bands. Through primarily nonradiative relaxation, these carriers form A-excitons that recombine radiatively and emit light at $\lambda_\mathrm{emi} = 610$~nm~\cite{Zhao2013Evolution}. Both the excitation and emission processes can be enhanced by coupling the emitter to a Mie void resonator. To determine the void depth $h$ and void diameter $d$ that maximize the PL signal, we perform three-dimensional electromagnetic simulations using the finite-element method in COMSOL Multiphysics (see Methods). The excitation process is modeled by a plane wave normally incident on a single void (Fig.~\ref{fig:2}a), while the emission process is modeled by an electric dipole oriented parallel to the sample surface and positioned laterally at the center of the void and vertically at its opening (Fig.~\ref{fig:2}b), matching the experimental configuration in which the encapsulated WS$_2$ monolayer rests on top of the WSe$_2$ voids.

Enhancement of the excitation process $\gamma_\mathrm{exc}$ is quantified through the intensity of the electric field component along the dipole oscillation direction $\mathbf{p}_\mathrm{dp}$ at the dipole position, i.e., $|\mathbf{E}(\mathbf{r}_\mathrm{dp}, \lambda_\mathrm{exc})\cdot\mathbf{p}_\mathrm{dp}|^2$. Enhancement of the emission process $\gamma_\mathrm{emi}$ is characterized by the total radiated power integrated over the solid angle corresponding to the numerical aperture (NA) of the objective lens used experimentally. The emitted power is calculated from the Poynting vector $\mathbf{S}$ of the dipole field. The total PL enhancement is obtained by normalizing the simulated and measured PL intensity from the void resonator to that from a flat unstructured WSe$_2$ region

\begin{align}\label{eq:PLenh}
PL_\text{enh}
&= \gamma_{\text{exc}}(\lambda_{\text{exc}}) \times \gamma_{\text{emi}}(\lambda_{\text{emi}}) \nonumber \\
&= \frac{\left|\mathbf{E}_{\text{void}}\left(\mathbf{r}_{\text{dp}},\lambda_{\text{exc}}\right)\cdot\mathbf{p}_\mathrm{dp}\right|^{2}}{\left|\mathbf{E}_{\text{flat}}\left(\mathbf{r}_{\text{dp}},\lambda_{\text{exc}}\right)\cdot\mathbf{p}_\mathrm{dp}\right|^{2}} \nonumber \\ 
    &\times 
    \frac{\int_{\textrm{NA}}\mathbf{S}_{\mathrm{void}}(\lambda_{\mathrm{emi}})\cdot\hat{\mathbf{r}}\,d\Omega}{\int_{\textrm{NA}}\mathbf{S}_\textrm{flat}(\lambda_{\text{emi}})\cdot\hat{\mathbf{r}}\,d\Omega},
\end{align}
where $\hat{\mathbf{r}}$ denotes the unit vector in the radial direction. 

%
Our simulations show that the PL enhancement strongly depends on the void dimensions. A void with a depth of 180~nm and a diameter of 520~nm yields the highest enhancement, reaching a 1700-fold increase in PL compared to unpatterned flat WSe$_2$ (Fig.~\ref{fig:2}c). Increasing the void depth moves the field maximum deeper into the cavity, thereby reducing the spatial overlap with the dipole at the surface and thus weakening their interaction (Supplementary~Section~1, Fig.~S1).

To disentangle the excitation and emission contributions, we separately compute their efficiencies as functions of the void dimensions (Fig.~\ref{fig:2}d-e). Both exhibit similar trends with changing void size, but their maxima occur under slightly different conditions. The excitation efficiency peaks for a void with a depth of $h=160$~nm and diameter of $d=540$~nm, while the emission efficiency reaches its maximum for a void with dimensions $h=200$~nm and $d=500$~nm. The difference in optimal dimensions arises because the excitation and emission enhancements occur at different wavelengths. 

The depth-diameter trends in the excitation and emission efficiencies can be understood from a constructive-interference model (see Supplementary~Section~2, Fig.~S2). In this model, light entering the void is approximated as a plane wave that propagates with an effective wavenumber $k_z=k_0\mathrm{Re}\left[n_\mathrm{eff}(d)\right]$, where $n_\mathrm{eff}$ is the effective index of the fundamental electric-dipolar mode of an air-WSe$_2$ cylindrical waveguide and depends on the diameter (see Fig.~S2). Maximal field enhancement at the void opening occurs when the downward-propagating wave, after reflection at the void bottom, returns in phase with the field at the aperture, which is satisfied when 
\begin{equation}\label{eq:constinter}
    2k_0\mathrm{Re}\left[n_\mathrm{eff}(d)\right]h + \phi = 2\pi.
\end{equation}
In Eq.~(\ref{eq:constinter}), $k_0=2\pi/\lambda$ is the free-space wavenumber and $\phi$ is the phase accumulated upon reflection at the void bottom, which is close to $\pi$ and nearly constant across all diameters (see Fig.~S1). It should be noted that the same condition also governs the directionality of the emission, since constructive interference between the upward-emitted and downward-reflected waves enhances radiation into the top ($+z$) direction. This model quantitatively captures the depth–diameter relation observed in the full-wave simulations of the excitation and emission efficiencies, as shown by the overlaid lines in Fig.~\ref{fig:2}d–e. The overall PL enhancement is therefore maximized when both efficiencies coincide, underscoring that the enhancement arises from a synergistic interplay between optimized field localization and directional far-field emission.

To quantify directionality, we compute the Poynting vector amplitude for a dipole placed on a void resonator and normalize it to that of a dipole on flat WSe$_2$ (Fig.~\ref{fig:2}f). The void-coupled configuration yields up to a 48-fold enhancement of the power flow, confirming that Mie voids efficiently direct emission into the out-of-plane direction. 
\begin{figure*}[!htb]
  \centering
  \includegraphics[width=\textwidth]{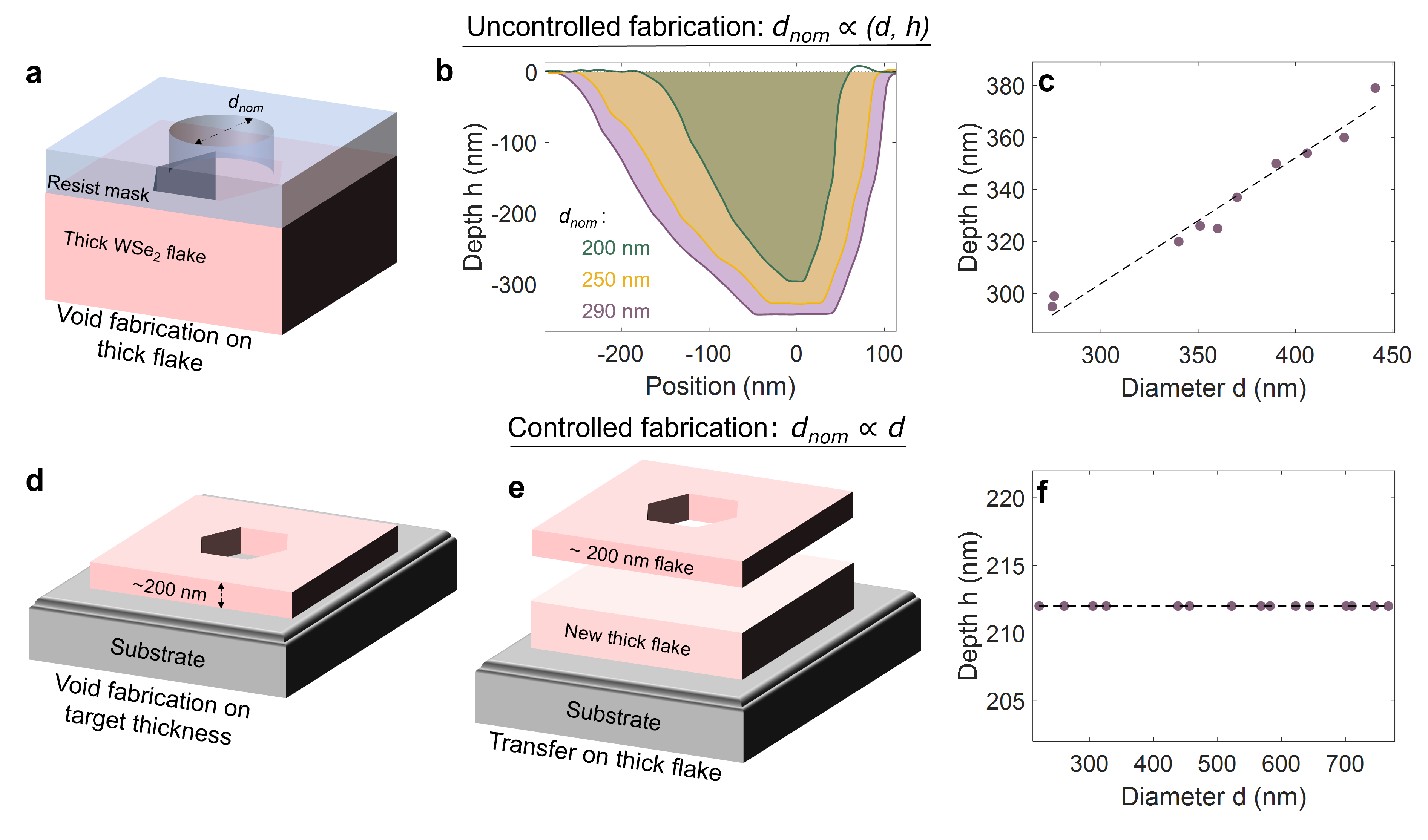}
  \caption{\textbf{Solving fabrication barrier.}
  \textbf{a,} Schematic of conventional lithographic fabrication of Mie voids in a WSe$_2$ flake (pink) with a thickness exceeding the targeted void depth. The etched void diameter can exceed the nominal diameter $d_\mathrm{nom}$ defined by the resist mask opening (blue) due to lateral etching.
  \textbf{b,} Atomic-force microscopy measurements of voids fabricated using identical etch times show that larger mask openings $d_\mathrm{nom}$ produces voids with both larger diameters $d$ and larger depths $h$. The surface is at $h=0$~nm.
  \textbf{c,} Measured void depth versus diameter for the conventional process, demonstrating a near-linear dependence on the void dimensions due to the mask-opening-dependent etch rate. 
  \textbf{d,e,} Mie voids fabricated using a two-step process to independently control the diameter and depth. A WSe$_2$ flake with a thickness matching the target depth ($\sim200$~nm) is lithographically patterned (\textbf{d}). The patterned flake is subsequently transferred onto an optically-thick WSe$_2$ that serves as a substrate to provide the bottom interface of the Mie void (\textbf{e}).
  \textbf{f,} Measured void depth versus diameter for the two-step fabrication process. The void diameter is controlled by the mask opening, while the depth remains constant at $h=212$~nm.}
  \label{fig:fab}
\end{figure*}

\subsection*{Fabricating voids with independently-controlled diameter and depth}
The electromagnetic simulations show that achieving maximal PL enhancement depends sensitively on the void dimensions (Fig.~\ref{fig:2}c). Consequently, it is crucial to independently control the void depth and void diameter in the fabrication process. However, when performing lithographic patterning at the nanoscale, the etch rate can depend on the nominal diameter $d_\mathrm{nom}$, which is dictated by the size of the opening in the resist mask~(Fig.~\ref{fig:fab}a). In particular, lateral and vertical etch rates differ, with smaller nominal diameters typically exhibiting slower vertical etching. In addition, for arrays of voids, the concentration of voids can further accelerate or retard etching, complicating precise control over void dimensions (Supplementary~Section~3, Fig.~S3)~\cite{Huff2021,Jensen2004,Wu2010,Carlstrom2006,Sridhar2012,Sarbajna2024}. This means that both the void diameter and void depth depend on the nominal diameter $d_\mathrm{nom}$ (Fig.~\ref{fig:fab}b). Larger mask openings yield deeper and wider voids. Figure~\ref{fig:fab}c quantifies the correlation between void depth and void diameter for samples etched for the same duration but with different nominal diameters. Similar observations have been reported in other materials and fabrication processes related to fabricating Mie voids, such as dry etching in GaAs~\cite{Ludescher2025} and focused ion beam milling in Si~\cite{Hentschel2023}. 

To decouple diameter and depth during fabrication, we develop a two-step process which leverages the layered nature of WSe$_2$ to independently control the void diameter via the mask size and the depth via mechanical exfoliation and transfer techniques~(see Methods). The first step is to exfoliate a WSe$_2$ flake with a thickness that matches the target void depth ($\sim200$~nm). As the void depth $h$ is dictated by this flake thickness, the opening in the resist mask now only controls the void diameter $d$~(Fig.~\ref{fig:fab}d). After lithographic patterning and etching, the second step is to transfer the thin flake with void resonators onto an optically-thick WSe$_2$ substrate to provide the bottom interface of the void resonator~(Fig.~\ref{fig:fab}e). This two-step process decouples the two geometric parameters of the void resonators, enabling controlled fabrication of voids with different diameters at a constant depth~(Fig.~\ref{fig:fab}f). The depth of all the voids is $h=212$~nm and the WSe$_2$ substrate beneath the resonators has a thickness of 2.7~\textmu m (see Methods). In the final step, we transfer an hBN-encapsulated monolayer of WS$_2$ onto the voids. The hBN layers have a thickness of $\sim5-7$~nm. The encapsulation protects the monolayer from environmental degradation, suppresses interlayer charge transfer and improves surface flatness~\cite{PhysRevB.98.115104, PhysRevX.7.021026}. We have conducted additional simulations to investigate whether the slight vertical displacement from the void opening due to the hBN encapsulation impacts the emission efficiency (see Supplementary~Section~4, Fig.~S4). These simulations show that the emission efficiency remains nearly constant even if the dipole is displaced by up to 20~nm above the surface, indicating that the thin hBN encapsulation has only a minor impact on the emission efficiency.

\subsection*{Photoluminescence enhancement from individual Mie voids}
\begin{figure}[tp]
    \centering
    \includegraphics[width=\columnwidth]{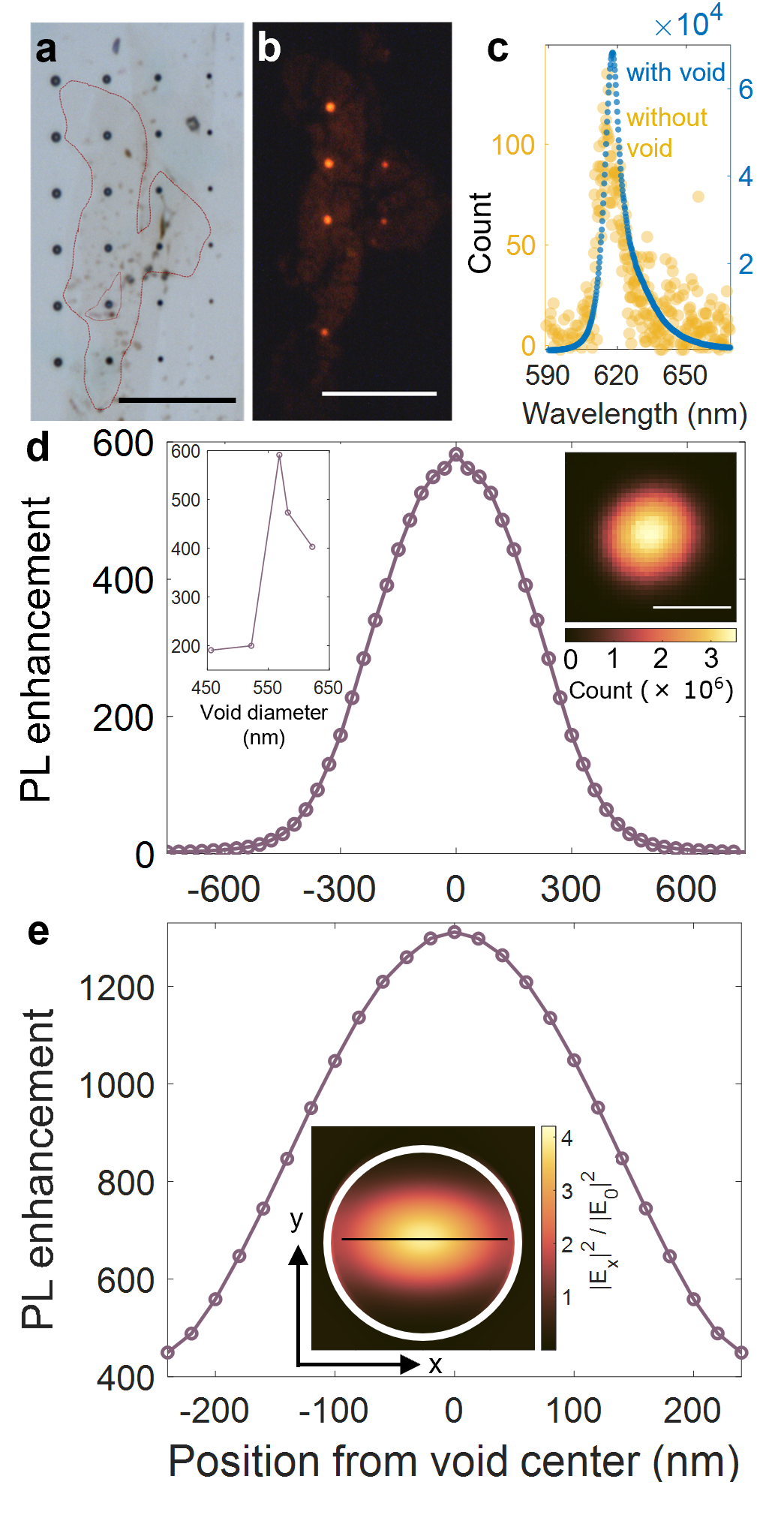}
    \caption{\textbf{Void-enabled photoluminescence enhancement.} \textbf{a,} Bright-field micrograph of the WS$_2$ monolayer on voids with different diameters. The red border indicates the monolayer. \textbf{b,} Corresponding optical image of PL emission showing much brighter emission from the voids relative to flat WSe$_2$. The scale bars for both the images are 10~\textmu m. \textbf{c,} Spectrally resolved PL signal with and without the void resonator. \textbf{d,} Experimental PL enhancement of monolayer-on-void structure as a function of the position from the void center. Top-right inset shows two-dimensional PL mapping of the same structure (scale bar: 500~nm). Top-left inset shows experimental peak PL enhancement of voids with varying diameters. \textbf{e,} Simulated PL enhancement as a function of the dipole position. The inset shows the normalized electric field profile at the void opening at an excitation wavelength of 532~nm. The black line indicates the axis along which the PL enhancement is calculated.}
    \label{fig:3}
\end{figure}
Figure~\ref{fig:3}a shows an optical micrograph of the encapsulated monolayer WS$_2$ stacked on voids of different diameters. PL imaging reveals that the void regions appear significantly brighter than the adjacent flat WSe$_2$ areas, indicating enhanced emission due to the void resonators (Fig.~\ref{fig:3}b). Spectrally-resolved PL measurements of the monolayer WS$_2$ on both the void resonator and the flat WSe$_2$ confirm that the observed emission arises from the A-exciton in monolayer WS$_2$ (Fig.~\ref{fig:3}c). The PL peaks around a wavelength of 620~nm for the monolayer on the void and 615~nm for the monolayer sitting on flat WSe$_2$, which are slightly red-shifted from the A-exciton wavelength of unencapsulated WS$_2$ ($\sim610$~nm). The hBN encapsulation decreases the exciton binding energy due to dielectric screening, thereby causing a red-shift in the PL emission~\cite{Uchiyama2019, GAO2023768, PhysRevLett.116.066803, PhysRevLett.113.026803}.

We perform PL mapping using a confocal setup (see Methods) to spatially resolve the emission across the void (Fig.~\ref{fig:3}d). The excitation beam has a spot diameter of approximately 1~\textmu m. The sample is mounted on a piezo stage, and the map is acquired by recording the PL signal in translation steps of 30~nm across the targeted area. The upper-right inset of Fig.~\ref{fig:3}d shows a zoomed-in PL map of a WS$_2$ monolayer stacked on top of a single void. The 1D intensity plot extracted from the map (Supplementary~Section~5, Fig.~S5) represents the PL enhancement. Benchmarking the void-assisted emission with that of emission from the flat WSe$_2$ substrate, we observe up to 600-fold PL enhancement at the void center, for a void with dimensions $d=568$~nm and $h=212$~nm. We further see that the PL emission peaks at the void center and decays almost symmetrically toward the rim.  
We extend these measurements to voids of different diameters and track the peak PL enhancement (Fig.~\ref{fig:3}d, top-left inset). We observe that the enhancement increases with increasing diameter until it reaches an optimal size, beyond which further increase of the void diameter causes the enhancement to drop. When the void diameter exceeds the optimal size, the excitation and emission wavelengths detune from the void resonance wavelength, resulting in a weakening of the field localization and, consequently, the PL enhancement. 

\begin{figure*}[!th]
  \centering
  \includegraphics[width=\textwidth]{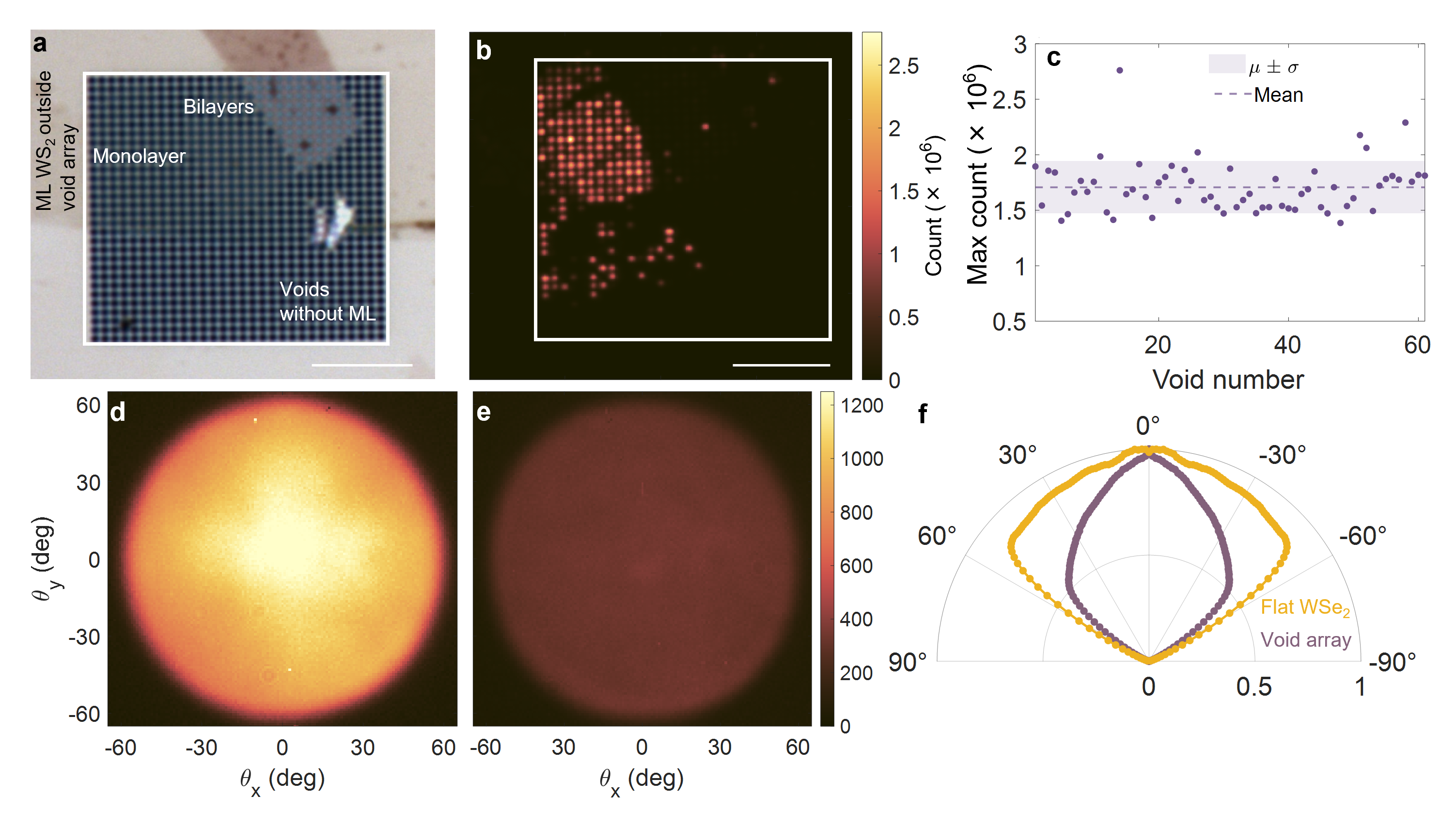}
  \caption{\textbf{Void-enabled directional emission.}
  \textbf{a,} Optical image of a WS$_2$ monolayer on a void array. Scale bar: 5~\textmu m.
  \textbf{b,} PL map of the same region, showing brighter emission from monolayer-on-void regions. Scale bar: 5~\textmu m.
  \textbf{c,} Maximum PL count of individual voids covered by monolayer WS$_2$. 
  The black dashed line indicates the sample mean ($\mu$) of the maximum PL intensity, and the shaded band shows the standard deviation ($\pm \sigma$) around the mean.
  \textbf{d,} Back-focal-plane image of a monolayer-on-array configuration.
  \textbf{e,} Back-focal-plane image of a monolayer-on-flat WSe$_2$ configuration.
  \textbf{f,} Azimuthally-averaged 1D angular PL plots (normalized) calculated from the back-focal-plane images in \textbf{d,e}.}
  \label{fig:4}
\end{figure*}

To understand the experimental results, we perform simulations of the spatially-resolved PL enhancement by translating a single dipole radially outward from the void center and calculate the emission efficiency at each position. By multiplying the spatially-resolved emission efficiencies with the normalized electric field intensities at the same positions extracted from the corresponding excitation efficiency simulations, we can evaluate the spatial dependence of the PL enhancement. Figure~\ref{fig:3}e shows the simulation result of the spatially-resolved PL enhancement for a void of identical dimensions to the experimental sample shown in Fig.~\ref{fig:3}d. The PL enhancement is maximal for a dipole positioned at the void center and decreases as the dipole shifts radially outwards, matching our experimental outcome. The simulated enhancement reaches around 1300$\times$, more than 2 times higher than the experimental value. The lower experimental enhancement is due to the diffraction-limited spot size of the illumination laser, which excites an ensemble of excitons distributed across the entire void. This is different from the simulation model, which considers only a single dipole at each position. Although the peak values differ, the spatial profile of the simulated PL enhancements is in good agreement with the experimental results. In addition, the spatial dependence of PL enhancement follows the electric field profile of the void resonance (Fig.~\ref{fig:3}e, inset). This confirms that the PL enhancement originates from the capability of the void resonance to enhance the local electric field and thereby enhance both excitation and emission efficiencies (see Supplementary~Section~4 and Fig.~S4 for spatially-resolved simulations of the excitation and emission efficiencies).

\subsection*{Directional emission using Mie voids}
A key contribution to the enhanced PL emission arises from the ability of Mie voids to direct the emitted light toward the collection objective. To investigate this directionality, we perform angle-resolved back-focal-plane imaging under wide-field illumination (see Methods). Because the collection area of the optical setup ($\approx 30$~\textmu m$^2$, see Supplementary~Section~6, Fig.~S6) is much larger than the area of a single void ($\approx 0.2$~\textmu m$^2$), measurements on individual voids are not feasible. We therefore fabricated void arrays with a 500~nm pitch and designed the void resonance wavelength to overlap spectrally with the WS$_2$ monolayer exciton wavelength (see Supplementary~Section~7 and Fig.~S7). This ensures that the collected signal predominantly originates from void-coupled regions. Figure~\ref{fig:4}a shows an optical micrograph of a WS$_2$ monolayer placed on such a void array (with $h=336$~nm and $d=406$~nm), and the corresponding PL map in Fig.~\ref{fig:4}b confirms enhanced emission localized above individual voids. The peak PL intensity shows only minor variations from void to void (see Methods), indicating uniform emission across the array (Fig.~\ref{fig:4}c).

Figures~\ref{fig:4}d,e display the raw back-focal-plane PL images from a WS$_2$ monolayer on a void array and on flat WSe$_2$, respectively. The axes ($\theta_x,\theta_y$) correspond to collection angles along the two lateral directions defined by the numerical aperture of the objective lens (\(\mathrm{NA}=0.90\)). The void array produces a bright central lobe whose intensity decreases toward the periphery, indicating that emission near the surface normal ($\theta \approx 0^\circ$) dominates. In contrast, the flat WSe$_2$ reference exhibits an almost angle-independent, Lambertian-like distribution. Because the azimuthal variations in Fig.~\ref{fig:4}d are relatively minor, we approximate the emission as radially symmetric and average the intensity over the azimuthal angle to obtain a one-dimensional polar intensity profile (using a similar process as described in Supplementary~Section~5).

To compare the void-coupled and planar emission, we account for the planar regions between neighboring voids that do not contribute to directionality. This is done by an area-weighted subtraction based on the planar fill factor, corresponding to the fraction of planar surface within the array period. The subtraction assumes that PL from different regions adds incoherently, which is appropriate for spontaneous emission. A detailed description of this procedure and its limited impact on the overall result are provided in Supplementary~Section~8 and Fig.~S8.

The resulting normalized angular profiles are shown in Fig.~\ref{fig:4}f. The void-coupled emission at normal incidence is approximately 83\% higher than that at angle of \ang{50}, corresponding to a forward-to-reference directionality of about 2.6~dB. For the flat WSe$_2$ reference, the emission changes by less than 10\% over the same angular range. The pronounced central beaming observed for the void arrays thus provides clear experimental evidence that Mie voids efficiently channel emission into the forward direction.

\section*{Discussion}
We have demonstrated that Mie voids provide a powerful and conceptually distinct route to control light emission at the nanoscale. By coupling an atomically thin WS$_2$ monolayer to Mie void resonators formed in WSe$_2$, we achieved simultaneous enhancement and directionality of photoluminescence within an all–van der Waals architecture. Electromagnetic simulations and optical measurements show that the PL enhancement arises from the synergistic interplay between strong field localization inside the air void and constructive interference between the emitter and the void resonance. Experimentally, this coupling leads to a 600-fold increase in PL intensity and pronounced beaming of light along the surface normal.  

The presented two-step fabrication method enables independent control of void diameter and depth, providing a pathway to engineer the optical response of van der Waals materials with nanoscale precision. Because Mie voids relax the need for low-loss dielectric materials, they can be realized in a wide range of high-index semiconductors and combined with diverse quantum emitters. This opens a new design space for compact, efficient, and directional light sources. Simulations further indicate that increasing the void depth shifts the field localization deeper inside the cavity, which may be attractive for integrating quantum dots or single molecules placed within the void. Stacking such a configuration with another layered material such as hBN could additionally shield the embedded emitters from the environment. These avenues remain open for exploration and may extend the Mie void platform toward quantum photonics based on hybrid van der Waals architectures.


\section*{Methods}

\subsection*{Fabrication}
The device fabrication proceeds through a sequential multi-step process (see Fig.~S8). WSe$_2$ flakes are mechanically exfoliated using 3M Scotch Magic Tape 810 from bulk WSe$_2$ crystals from HQ Graphene onto a Si substrate and inspected to identify flakes matching the desired thickness and surface quality for subsequent void patterning. Then we spin-coat electron-beam resist to define the etch mask. A \SI{4}{wt\percent} PMMA (996\,k) solution in anisole is spun at \SI{2000}{rpm} for \SI{60}{s} and baked at \SI{170}{\celsius} for \SI{300}{s}. 
Circular mask openings are then patterned by electron-beam lithography using a \SI{30}{kV} Raith eLINE Plus with a \SI{30}{\micro\metre} aperture and dose of \SI{230}{\micro C\per cm^{2}}. The sample is developed in IPA:H$_2$O (3:1, \SI{60}{s}) followed by an IPA rinse (\SI{30}{s}). Then the voids are fabricated into WSe$_2$ by reactive-ion etching. First we use an initial O$_2$ plasma descum to remove residual resist, followed by an SF$_6$ plasma etch to define the voids. We use two parameter sets: (i) O$_2$ plasma clean (\SI{9.15}{mTorr}, \SI{30}{W}, \SI{30}{sccm}, \SI{10}{s}), followed by an SF$_6$ etch (\SI{9.15}{mTorr}, \SI{30}{W}, \SI{30}{sccm}, \SI{33}{s}); and (ii) O$_2$ plasma descum (\SI{10}{mTorr}, \SI{30}{W}, \SI{40}{sccm}, \SI{10}{s}), followed by an SF$_6$ etch (\SI{10}{mTorr}, \SI{30}{W}, \SI{30}{sccm}, \SI{150}{s}). The resist is stripped in acetone and the sample is rinsed sequentially in IPA and DI water, leaving open voids.

The next step is to prepare thick WSe$_2$ flakes that serve as the bottom WSe$_2$ substrate underneath the resonators. Thick WSe$_2$ flakes from bulk crystals (HQ Graphene) are exfoliated using thermal-release tape on a substrate. Applying heat of roughly \SI{100}{\celsius} disengages the tape adhesive, allowing us to transfer pristine flakes (up to a few micrometres thick). The etched WSe$_2$ flakes containing the voids are transferred onto such a thick WSe$_2$ flake. We assemble the stack using a dry-transfer method with a 10\% polycarbonate (PC) film on a polydimethylsiloxane (PDMS) stamp (HQ Graphene transfer system, product code HQ2D MOT). The void-patterned thin flake is picked up at \SI{110}{\celsius}. For placement on the thick WSe$_2$ flake, the stage is heated to \SI{180}{\celsius} to melt the PC and release the stack. Residual PC is removed by immersion in chloroform for \SI{10}{\minute}.

For large-area monolayers, we employ Au-assisted exfoliation combined with thermal release tape (TRT). Thin flakes are first mechanically exfoliated using Scotch tape (3M Scotch Magic Tape 810). A 50~nm Au film is then thermally evaporated onto the tape under high vacuum ($6 \times 10^{-6}$~mbar) using an Oerlikon UNIVEX~250 evaporation chamber. The Au thickness is determined from a deposition rate of $0.3~\text{\AA}\,\text{s}^{-1}$, and the Au layer serves as an adhesion layer for the subsequent transfer process. The Au-coated exfoliation tape is then laminated onto a thermal release tape such that the flakes remain on top while the Au layer is positioned underneath. This Au/TRT stack is subsequently brought into contact with the target substrate and gently pressed to ensure uniform adhesion of the flakes. The sample is then heated to activate the thermal release tape, allowing clean detachment of the tape while leaving the flakes adhered to the substrate with residual Au. The remaining Au is removed by etching in KI solution (or aqua regia), followed by rinsing in deionized (DI) water. No additional cleaning steps are performed to preserve the pick-up yield. The high surface energy of Au and its uniform adhesion selectively ``pull'' the outermost TMD layer, enabling millimetre-scale monolayer exfoliation.
%

The WS$_2$ monolayer is encapsulated in hBN and the resulting hBN/WS$_2$/hBN stack is transferred onto the void device to complete the heterostructure. We use the same PC/PDMS dry-transfer approach as previously described. Thin TMDC flakes can typically be picked up at \SIrange{90}{100}{\celsius}, while for monolayer crystals we find that increasing the temperature to roughly \SI{120}{\celsius} yields more reliable adhesion to PC.

\subsection*{Photoluminescence measurements}
We use a custom-built confocal microscope to record high‐resolution PL maps and spectra. Excitation is provided by a 532~nm solid‐state laser (Cobolt, Hübner GmbH), whose power is varied via a rotating half‐wave plate. Both excitation and collection are performed through a diffraction‐limited Nikon TU Plan Fluor 100×, NA = 0.95 EPI D objective. The laser power for the measurement is around 30~\textmu W. Confocal detection is realized by coupling the collected PL into a single‐mode optical fiber, and sample scanning is achieved with a three-axis piezoelectric stage. In the detection path, a 550~nm dichroic mirror and a 532~nm notch filter separate excitation light from the emitted PL. PL is recorded using an avalanche photodiode (APD) single-photon detector. To maximize collection efficiency, we first scan the sample position along the z-axis and record the photon counts at different distances between the stage and the objective. We then identify the z-position that yields the highest photon count and perform the PL mapping at this focus ~(see Fig.~S9). For spectrally resolved measurements, the PL output is diverted via a fiber‐coupled beamsplitter into a spectrometer equipped with a piezo‐cooled CCD camera (Andor-Solis SR-303i, 150 grooves/mm grating with blaze at 800 nm).
To determine the PL peak intensity distribution shown in Fig.~\ref{fig:4}c, we include only those voids whose maximum PL intensity is at least 50\% of the peak intensity of the brightest void in the array.

\subsection*{Angle-resolved photoluminescence measurements and bright-field microscopy}
Bright-field micrographs were obtained using a Nikon Eclipse LV100ND microscope under white-light illumination (Thorlabs OSL2). For the PL images we used a CoolLED pE800 LED system with a broadband illumination ranging 525-550~nm. We only collect the emission from the A-exciton using a band-pass filter (Semrock, ranging 590-650~nm) and a dichroic mirror (FF570-Di01).

The back-focal-plane PL imaging setup, consisting of three free-space lenses, was integrated with the above mentioned optical microscope and an Andor Kymera 328i spectrograph to capture angle-resolved images in $k$-space. To obtain the emitted signal we employed the same broadband excitation ranging 525-555~nm mentioned before and collected the emission between 590-650~nm with a high-NA (0.90) objective to access the widest angular range. An iris was placed at the intermediate image plane to ensure that we collect emission only from the region of interest, while a relay lens reimaged the back focal plane of the objective onto the spectrograph. This configuration converted the real-space image into its Fourier-plane representation, providing direct access to the angular distribution of the emission. The emission angles were retrieved from the recorded pixel positions using the relation $\sin\theta = (r/R_{\text{max}})\,\mathrm{NA}/n$, where $r$ is the radial pixel distance from the optical axis, $R_{\text{max}}$ is the pupil radius in pixel, $\mathrm{NA}$ is the numerical aperture of the objective, and $n$ is the refractive index of the collection medium (1 for air). The optical setup is similar to that used in our earlier work~\cite{Danielsen2025}.

\subsection*{Electromagnetic simulations}
Three-dimensional electromagnetic simulations are conducted using COMSOL Multiphysics, which solves Maxwell's equations using the finite element method. The simulation model consists of a WSe$_2$-air interface with a cylindrical void etched into the WSe$_2$ surface (see Fig.~S9). The coordinate system is defined with $z$ as the surface normal and $(x,y)$ as the in-plane coordinates. Both the linear polarization of the incident excitation plane-wave field and the dipole moment of the emitting dipole are along $x$ (see Fig.~S9). The complex anisotropic refractive index of WSe$_2$ is taken from Ref.~\cite{munkhbat2022optical}. Domain terminations were as follows: top and lateral boundaries (for air domain) used perfectly matched layers to absorb outgoing fields, while the air-WSe$_2$ interface is modeled using an impedance boundary condition. To evaluate the emission efficiency, the Poynting vector was integrated on a plane with an area dictated by the numerical aperture of the objective lens ($\mathrm{NA}=0.9$), such that only angles $\theta \le \arcsin(\mathrm{NA})$ are collected. The integration plane is placed at a vertical distance of 1.5~\textmu m from the void surface, so that the recorded fields are in the far-field regime.


\bibliography{main}

@article{Lepeshov2019,
  author  = {Sergey Lepeshov and Alex Krasnok and Andrea Al{\`u}},
  title   = {Enhanced Excitation and Emission from {2D} Transition Metal Dichalcogenides with All-Dielectric Nanoantennas},
  journal = {Nanotechnology},
  year    = {2019},
  volume  = {30},
  number  = {25},
  pages   = {254004},
  doi     = {10.1088/1361-6528/ab0daf},
  url     = {https://iopscience.iop.org/article/10.1088/1361-6528/ab0daf},
}

@article{Zhang2015,
  author  = {Shouren Zhang and Ruibin Jiang and Ya-Ming Xie and Qifeng Ruan and Baocheng Yang and Jianfang Wang and Hai-Qing Lin},
  title   = {Colloidal Moderate-Refractive-Index {Cu}$_2${O} Nanospheres as Visible-Region Nanoantennas with Electromagnetic Resonance and Directional Light-Scattering Properties},
  journal = {Adv. Mater.},
  year    = {2015},
  volume  = {27},
  number  = {45},
  pages   = {7432--7439},
  doi     = {10.1002/adma.201502917},
  url     = {https://onlinelibrary.wiley.com/doi/10.1002/adma.201502917},
}

@article{Ozawa2025,
  author  = {Keisuke Ozawa and Hiroshi Sugimoto and Daisuke Shima and Tatsuki Hinamoto and Mojtaba Karimi Habil and Yan Joe Lee and Søren Raza and Keisuke Imaeda and Kosei Ueno and Mark L. Brongersma and Minoru Fujii},
  title   = {Routing Light Emission from Monolayer {MoS}$_2$ by {Mie} Resonances of Crystalline Silicon Nanospheres},
  journal = {ACS Appl. Opt. Mater.},
  year    = {2025},
  volume  = {3},
  pages   = {375--382},
  doi     = {10.1021/acsaom.4c00495},
  url     = {https://pubs.acs.org/doi/10.1021/acsaom.4c00495}
}

@article{Vaskin2018,
   author = {Aleksandr Vaskin and Justus Bohn and Katie E. Chong and Tobias Bucher and Matthias Zilk and Duk-Yong Choi and Dragomir N. Neshev and Yuri S. Kivshar and Thomas Pertsch and Isabelle Staude},
   doi = {10.1021/acsphotonics.7b01375},
   journal = {ACS Photonics},
   month = {4},
   pages = {1359-1364},
   title = {Directional and Spectral Shaping of Light Emission with {Mie}-Resonant Silicon Nanoantenna Arrays},
   volume = {5},
   url = {https://pubs.acs.org/doi/10.1021/acsphotonics.7b01375},
   year = {2018}
}

@article{Fang2021,
   author = {Jie Fang and Mingsong Wang and Kan Yao and Tianyi Zhang and Alex Krasnok and Taizhi Jiang and Junho Choi and Ethan Kahn and Brian A. Korgel and Mauricio Terrones and Xiaoqin Li and Andrea Alù and Yuebing Zheng},
   doi = {10.1002/adma.202007236},
   journal = {Adv. Mater.},
   pages = {2007236},
   title = {Directional Modulation of Exciton Emission Using Single Dielectric Nanospheres},
   volume = {33},
   url = {https://advanced.onlinelibrary.wiley.com/doi/10.1002/adma.202007236},
   year = {2021}
}

@article{Rolly2012,
   author = {Brice Rolly and Brian Stout and Nicolas Bonod},
   doi = {10.1364/OE.20.020376},
   issn = {1094-4087},
   issue = {18},
   journal = {Opt. Express},
   month = {8},
   pages = {20376},
   title = {Boosting the directivity of optical antennas with magnetic and electric dipolar resonant particles},
   volume = {20},
   url = {https://opg.optica.org/oe/abstract.cfm?uri=oe-20-18-20376},
   year = {2012}
}

@article{Rolly2013,
   author = {Brice Rolly and Jean Michel Geffrin and Redha Abdeddaim and Brian Stout and Nicolas Bonod},
   doi = {10.1038/srep03063},
   issn = {20452322},
   journal = {Sci. Rep.},
   pages = {3063},
   title = {Controllable emission of a dipolar source coupled with a magneto-dielectric resonant subwavelength scatterer},
   volume = {3},
   year = {2013}
}

@article{Novotny2011,
  author    = {Lukas Novotny and Niek F. van Hulst},
  title     = {Antennas for light},
  journal   = {Nat. Photon.},
  volume    = {5},
  number    = {2},
  pages     = {83--90},
  year      = {2011},
  doi       = {10.1038/nphoton.2010.237}
}

@article{Curto2010,
  author    = {Alberto G. Curto and Giorgio Volpe and Tim H. Taminiau and Mark P. Kreuzer and Romain Quidant and Niek F. van Hulst},
  title     = {Unidirectional emission of a quantum dot coupled to a nanoantenna},
  journal   = {Science},
  volume    = {329},
  number    = {5994},
  pages     = {930--933},
  year      = {2010},
  doi       = {10.1126/science.1191922}
}

@article{Coenen2014,
  author    = {Toon Coenen and Felipe Bernal Arango and A. Femius Koenderink and Albert Polman},
  title     = {Directional emission from a single plasmonic scatterer},
  journal   = {Nat. Commun.},
  volume    = {5},
  pages     = {3250},
  year      = {2014},
  doi       = {10.1038/ncomms4250}
}

@article{Hentschel2023,
  author    = {Mario Hentschel and Kirill Koshelev and Florian Sterl and Steffen Both and Julian Karst and Lida Shamsafar and Thomas Weiss and Yuri Kivshar and Harald Giessen},
  title     = {Dielectric {Mie} voids: confining light in air},
  journal   = {Light Sci. Appl.},
  volume    = {12},
  pages     = {3},
  year      = {2023},
  doi       = {10.1038/s41377-022-01015-z}
}

@article{Sarbajna2024,
  author    = {Avishek Sarbajna and Dorte Rubæk Danielsen and Laura Nevenka Casses and Nicolas Stenger and Peter Bøggild and Søren Raza},
  title     = {Encapsulated Void Resonators in {van der Waals} Heterostructures},
  journal   = {Laser Photonics Rev.},
  volume    = {19},
  number    = {3},
  pages     = {2401215},
  year      = {2024},
  doi       = {10.1002/lpor.202401215}
}

@article{Li2020,
  author    = {Nannan Li and Yunhe Lai and Shiu Hei Lam and Haoyuan Bai and Lei Shao and Jianfang Wang},
  title     = {Directional Control of Light with Nanoantennas},
  journal   = {Adv. Opt. Mater.},
  volume    = {9},
  number    = {1},
  pages     = {2001081},
  year      = {2020},
  doi       = {10.1002/adom.202001081}
}

@article{Cihan2018,
  author    = {Ahmet Fatih Cihan and Alberto G. Curto and Søren Raza and Pieter G. Kik and Mark L. Brongersma},
  title     = {Silicon {Mie} resonators for highly directional light emission from monolayer {MoS}$_2$},
  journal   = {Nat. Photon.},
  volume    = {12},
  number    = {5},
  pages     = {284--290},
  year      = {2018},
  doi       = {10.1038/s41566-018-0155-y}
}

@article{Staude2013,
  author    = {Staude, Isabelle and Miroshnichenko, Andrey E. and Decker, Manuel and Fofang, Nche T. and Liu, Sheng and Gonzales, Edward and Dominguez, Jason and Luk, Ting Shan and Neshev, Dragomir N. and Brener, Igal and Kivshar, Yuri},
  title     = {Tailoring Directional Scattering through Magnetic and Electric Resonances in Subwavelength Silicon Nanodisks},
  journal   = {ACS Nano},
  year      = {2013},
  volume    = {7},
  number    = {9},
  pages     = {7824--7832},
  doi       = {10.1021/nn402736f},
  url       = {https://doi.org/10.1021/nn402736f}
}

@article{Person2013,
  author    = {Person, S. and Jain, M. and Lapin, Z. and S{\'a}enz, J. J. and Wicks, G. and Novotny, L.},
  title     = {Demonstration of Zero Optical Backscattering from Single Nanoparticles},
  journal   = {Nano Lett.},
  volume    = {13},
  number    = {4},
  pages     = {1806--1810},
  year      = {2013},
  doi       = {10.1021/nl3041794}
}

@article{Hamidi2025,
  author    = {Hamidi, Masoud and Koshelev, Kirill and Gladyshev, Sergei and Can{\'o}s Valero, Adri{\`a} and Hentschel, Mario and Giessen, Harald and Kivshar, Yuri and Weiss, Thomas},
  title     = {Quasi-{B}abinet principle in dielectric resonators and {Mie} voids},
  journal   = {Phys. Rev. Res.},
  volume    = {7},
  number    = {1},
  pages     = {013136},
  year      = {2025},
  month     = {feb},
  publisher = {American Physical Society}
}

@article{PhysRevX.7.021026,
  title = {Excitonic Linewidth Approaching the Homogeneous Limit in {MoS}$_2$-Based van der {W}aals Heterostructures},
  author = {Cadiz, F. and Courtade, E. and Robert, C. and Wang, G. and Shen, Y. and Cai, H. and Taniguchi, T. and Watanabe, K. and Carrere, H. and Lagarde, D. and Manca, M. and Amand, T. and Renucci, P. and Tongay, S. and Marie, X. and Urbaszek, B.},
  journal = {Phys. Rev. X},
  volume = {7},
  issue = {2},
  pages = {021026},
  numpages = {12},
  year = {2017},
  month = {May},
  publisher = {American Physical Society},
  doi = {10.1103/PhysRevX.7.021026},
  url = {https://link.aps.org/doi/10.1103/PhysRevX.7.021026}
}

@article{PhysRevB.98.115104,
  title = {Interlayer excitons in transition metal dichalcogenide heterostructures},
  author = {Van der Donck, M. and Peeters, F. M.},
  journal = {Phys. Rev. B},
  volume = {98},
  issue = {11},
  pages = {115104},
  numpages = {10},
  year = {2018},
  month = {Sep},
  publisher = {American Physical Society},
  doi = {10.1103/PhysRevB.98.115104},
  url = {https://link.aps.org/doi/10.1103/PhysRevB.98.115104}
}

@article{Danielsen2025,
  author = {Dorte Rubæk Danielsen and Nolan Lassaline and Sander Jæger Linde and Magnus Vejby Nielsen and Xavier Zambrana-Puyalto and Avishek Sarbajna and Duc Hieu Nguyen and Timothy J. Booth and Nicolas Leitherer-Stenger and Søren Raza},
  title = {Fourier-Tailored Light–Matter Coupling in van der {W}aals Heterostructures},
  journal = {ACS Nano},
  volume = {19},
  number = {22},
  pages = {20645--20654},
  year = {2025},
  doi = {10.1021/acsnano.5c02025},
  url = {https://doi.org/10.1021/acsnano.5c02025},
  publisher = {American Chemical Society},
  issn = {1936-0851}
}

@article{munkhbat2022optical,
  title={Optical Constants of Several Multilayer Transition Metal Dichalcogenides Measured by Spectroscopic Ellipsometry in the 300–1700 nm Range: High Index, Anisotropy, and Hyperbolicity},
  author={Munkhbat, Battulga and Wróbel, Piotr and Antosiewicz, Tomasz J. and Shegai, Timur O.},
  journal={ACS Photonics},
  volume={9},
  number={7},
  pages={2398--2407},
  year={2022},
  publisher={American Chemical Society},
  doi={10.1021/acsphotonics.2c00433}
}

@article{Peter2017Directional,
  author    = {Peter, Manuel and Hildebrandt, Andre and Schlickriede, Christian and Gharib, Kimia and Zentgraf, Thomas and Förstner, Jens and Linden, Stefan},
  title     = {Directional Emission from Dielectric Leaky-Wave Nanoantennas},
  journal   = {Nano Lett.},
  volume    = {17},
  number    = {7},
  pages     = {4178--4183},
  year      = {2017},
  month     = {jul},
  doi       = {10.1021/acs.nanolett.7b00966},
  publisher = {American Chemical Society},
  issn      = {1530-6984},
  url       = {https://doi.org/10.1021/acs.nanolett.7b00966}
}

@article{Zhao2013Evolution,
  author    = {Zhao, Weijie and Ghorannevis, Zohreh and Chu, Leiqiang and Toh, Minglin and Kloc, Christian and Tan, Ping-Heng and Eda, Goki},
  title     = {Evolution of Electronic Structure in Atomically Thin Sheets of {WS$_2$} and {WSe$_2$}},
  journal   = {ACS Nano},
  volume    = {7},
  number    = {1},
  pages     = {791--797},
  year      = {2013},
  month     = {jan},
  doi       = {10.1021/nn305275h},
  publisher = {American Chemical Society},
  issn      = {1936-0851},
  url       = {https://doi.org/10.1021/nn305275h}
}

@article{Uchiyama2019,
  title     = {Momentum-Forbidden Dark Excitons in {hBN}-Encapsulated Monolayer {MoS}$_2$},
  author    = {Uchiyama, Yosuke and Kutana, Alex and Watanabe, Kenji and Taniguchi, Takashi and Kojima, Kana and Endo, Takahiko and Miyata, Yasumitsu and Shinohara, Hisanori and Kitaura, Ryo},
  year      = {2019},
  month     = {Jul},
  date      = {2019-07-19},
  journal   = {npj 2D Mater. Appl.},
  volume    = {3},
  number    = {1},
  pages     = {26}
}

@article{GAO2023768,
title = {Photoluminescence manipulation in two-dimensional transition metal dichalcogenides},
journal = {J. Materiomics},
volume = {9},
number = {4},
pages = {768-786},
year = {2023},
month = {July},
issn = {2352-8478},
doi = {https://doi.org/10.1016/j.jmat.2023.02.005},
url = {https://www.sciencedirect.com/science/article/pii/S2352847823000369},
author = {Minglang Gao and Lingxiao Yu and Qian Lv and Feiyu Kang and Zheng-Hong Huang and Ruitao Lv}
}

@article{PhysRevLett.116.066803,
  title = {Exciton Band Structure in Two-Dimensional Materials},
  author = {Cudazzo, Pierluigi and Sponza, Lorenzo and Giorgetti, Christine and Reining, Lucia and Sottile, Francesco and Gatti, Matteo},
  journal = {Phys. Rev. Lett.},
  volume = {116},
  issue = {6},
  pages = {066803},
  numpages = {6},
  year = {2016},
  month = {Feb},
  publisher = {American Physical Society},
  doi = {10.1103/PhysRevLett.116.066803},
  url = {https://link.aps.org/doi/10.1103/PhysRevLett.116.066803}
}

@article{PhysRevLett.113.026803,
  title = {Tightly Bound Excitons in Monolayer {WSe}$_2$},
  author = {He, Keliang and Kumar, Nardeep and Zhao, Liang and Wang, Zefang and Mak, Kin Fai and Zhao, Hui and Shan, Jie},
  journal = {Phys. Rev. Lett.},
  volume = {113},
  issue = {2},
  pages = {026803},
  numpages = {5},
  year = {2014},
  month = {Jul},
  publisher = {American Physical Society},
  doi = {10.1103/PhysRevLett.113.026803},
  url = {https://link.aps.org/doi/10.1103/PhysRevLett.113.026803}
}

@article{Ludescher2025,
    author = {Ludescher, D. and Wesemann, L. and Schwab, J. and Karst, J. and Sulejman, S. B. and Ubl, M. and Clarke, B. O. and Roberts, A. and Giessen, H. and Hentschel, M.},
    title = {Optical sieve for nanoplastic detection, sizing and counting},
    journal = {Nat. Photon.},
    year = {2025},
    month = {10},
    day = {01},
    volume = {19},
    number = {10},
    pages = {1138--1145},
    issn = {1749-4893},
    doi = {10.1038/s41566-025-01733-x},
    url = {https://doi.org/10.1038/s41566-025-01733-x}
}

@article{Huff2021,
  author  = {Huff, Michael},
  title   = {Recent Advances in Reactive Ion Etching and Applications of High-Aspect-Ratio Microfabrication},
  journal = {Micromachines},
  year    = {2021},
  volume  = {12},
  number  = {8},
  pages   = {991},
  doi     = {10.3390/mi12080991},
  url     = {https://doi.org/10.3390/mi12080991}
}

@inproceedings{Jensen2004,
  author    = {Jensen, S{\o}ren and Hansen, Ole},
  title     = {Characterization of the Microloading Effect in Deep Reactive Ion Etching of Silicon},
  booktitle = {Proceedings of SPIE — Micromachining and Microfabrication},
  year      = {2004},
  volume    = {5342},
  pages     = {111--118},
  publisher = {SPIE},
  address   = {Bellingham, WA},
  doi       = {10.1117/12.524461},
  url       = {https://doi.org/10.1117/12.524461}
}

@article{Wu2010,
  author  = {Wu, B. and Kumar, A. and Pamarthy, S.},
  title   = {High Aspect Ratio Silicon Etch: A Review},
  journal = {J. Appl. Phys.},
  year    = {2010},
  volume  = {108},
  number  = {5},
  pages   = {051101},
  doi     = {10.1063/1.3474652},
  url     = {https://doi.org/10.1063/1.3474652}
}

@article{Carlstrom2006,
  author  = {Carlstr{\"o}m, C. F. V. and van der Heijden, R. and Karouta, F. and van der Heijden, R. W. and Salemink, H. W. M. and van der Drift, E.},
  title   = {{Cl}\textsubscript{2}/{O}\textsubscript{2}-Inductively Coupled Plasma Etching of Deep Hole-Type Photonic Crystals in {InP}},
  journal = {J. Vac. Sci. Technol. B},
  year    = {2006},
  volume  = {24},
  number  = {1},
  pages   = {L6--L8},
  doi     = {10.1116/1.2151915},
  url     = {https://doi.org/10.1116/1.2151915}
}

@article{Sridhar2012,
  author  = {Sridhar, Manoj and Maurya, Devendra K. and Friend, James R. and Yeo, Leslie Y.},
  title   = {Focused Ion Beam Milling of Microchannels in Lithium Niobate},
  journal = {Biomicrofluidics},
  year    = {2012},
  volume  = {6},
  number  = {1},
  pages   = {012819},
  doi     = {10.1063/1.3673260},
  url     = {https://doi.org/10.1063/1.3673260}
}

@article{Chaubey2021,
author = {Chaubey, Shailendra K. and A, Gokul M. and Paul, Diptabrata and Tiwari, Sunny and Rahman, Atikur and Kumar, G. V. Pavan},
title = {Directional Emission from Tungsten Disulfide Monolayer Coupled to Plasmonic Nanowire-on-Mirror Cavity},
journal = {Adv. Photonics Res.},
volume = {2},
number = {6},
pages = {2100002},
year = {2021}
}

@article{
Kuznetsov2016,
author = {Arseniy I. Kuznetsov  and Andrey E. Miroshnichenko  and Mark L. Brongersma  and Yuri S. Kivshar  and Boris Luk’yanchuk },
title = {Optically resonant dielectric nanostructures},
journal = {Science},
volume = {354},
number = {6314},
pages = {aag2472},
year = {2016},
doi = {10.1126/science.aag2472}}

@article{Fu2013,
  author = {Fu, Yuan Hsing and Kuznetsov, Arseniy I. and Miroshnichenko, Andrey E. and Yu, Ye Feng and Luk'yanchuk, Boris},
  title = {Directional visible light scattering by silicon nanoparticles},
  journal = {Nat. Commun.},
  year = {2013},
  volume = {4},
  number = {1},
  pages = {1527},
  month = {feb},
  doi = {10.1038/ncomms2538},
  url = {https://doi.org/10.1038/ncomms2538},
  issn = {2041-1723}
}

@article{Tong2013,
  author = {Tong, L. and Pakizeh, T. and Feuz, L. and Dmitriev, A.},
  title = {Highly directional bottom-up {3D} nanoantenna for visible light},
  journal = {Sci. Rep.},
  volume = {3},
  number = {1},
  pages = {2311},
  year = {2013},
  month = {jul},
  day = {30},
  issn = {2045-2322},
  doi = {10.1038/srep02311},
  url = {https://doi.org/10.1038/srep02311}
}

@article{Shegai2011,
  author = {Shegai, Timur and Miljković, Vladimir D. and Bao, Kui and Xu, Hongxing and Nordlander, Peter and Johansson, Peter and Käll, Mikael},
  title = {Unidirectional Broadband Light Emission from Supported Plasmonic Nanowires},
  journal = {Nano Lett.},
  year = {2011},
  volume = {11},
  number = {2},
  pages = {706--711},
  month = {feb},
  day = {9},
  publisher = {American Chemical Society},
  issn = {1530-6984},
  doi = {10.1021/nl103834y},
  url = {https://doi.org/10.1021/nl103834y}
}

@article{Palacios2017,
  author = {Palacios, Edgar and Park, Spencer and Lauhon, Lincoln and Aydin, Koray},
  title = {Identifying Excitation and Emission Rate Contributions to Plasmon-Enhanced Photoluminescence from Monolayer {MoS$_2$} Using a Tapered Gold Nanoantenna},
  journal = {ACS Photonics},
  year = {2017},
  volume = {4},
  number = {7},
  pages = {1602--1606},
  month = {jul},
  day = {19},
  publisher = {American Chemical Society},
  doi = {10.1021/acsphotonics.7b00226},
  url = {https://doi.org/10.1021/acsphotonics.7b00226}
}

@article{Bidault2019,
    author = {Bidault, Sébastien and Mivelle, Mathieu and Bonod, Nicolas},
    title = {Dielectric nanoantennas to manipulate solid-state light
          emission},
    journal = {J. Appl. Phys.},
    volume = {126},
    number = {9},
    pages = {094104},
    year = {2019},
    month = {09},
    issn = {0021-8979},
    doi = {10.1063/1.5108641},
    url = {https://doi.org/10.1063/1.5108641}
}

@article{Murai2020,
author = {Murai, Shunsuke and Castellanos, Gabriel W. and Raziman, T. V. and Curto, Alberto G. and Rivas, Jaime Gómez},
title = {Enhanced Light Emission by Magnetic and Electric Resonances in Dielectric Metasurfaces},
journal = {Adv. Opt. Mater.},
volume = {8},
number = {16},
pages = {1902024},
doi = {https://doi.org/10.1002/adom.201902024},
url = {https://advanced.onlinelibrary.wiley.com/doi/abs/10.1002/adom.201902024},
year = {2020}
}

@article{Bucher2019,
  author = {Bucher, Tobias and Vaskin, Aleksandr and Mupparapu, Rajeshkumar and Löchner, Franz J. F. and George, Antony and Chong, Katie E. and Fasold, Stefan and Neumann, Christof and Choi, Duk-Yong and Eilenberger, Falk and Setzpfandt, Frank and Kivshar, Yuri S. and Pertsch, Thomas and Turchanin, Andrey and Staude, Isabelle},
  title = {Tailoring Photoluminescence from {MoS$_2$} Monolayers by {Mie}-Resonant Metasurfaces},
  journal = {ACS Photonics},
  year = {2019},
  volume = {6},
  number = {4},
  pages = {1002--1009},
  month = {apr},
  day = {17},
  publisher = {American Chemical Society},
  doi = {10.1021/acsphotonics.8b01771},
  url = {https://doi.org/10.1021/acsphotonics.8b01771}
}

@article{Huang2022,
  author = {Huang, Lujun and Krasnok, Alex and Alú, Andrea and Yu, Yiling and Neshev, Dragomir and Miroshnichenko, Andrey E.},
  title = {Enhanced light-matter interaction in two-dimensional transition metal dichalcogenides},
  journal = {Rep. Prog. Phys.},
  year = {2022},
  volume = {85},
  number = {4},
  pages = {046401},
  month = {mar},
  day = {8},
  doi = {10.1088/1361-6633/ac45f9}
}

@article{Sortino2019,
  author = {Sortino, L. and Zotev, P. G. and Mignuzzi, S. and Cambiasso, J. and Schmidt, D. and Genco, A. and Aßmann, M. and Bayer, M. and Maier, S. A. and Sapienza, R. and Tartakovskii, A. I.},
  title = {Enhanced light-matter interaction in an atomically thin semiconductor coupled with dielectric nano-antennas},
  journal = {Nat. Commun.},
  year = {2019},
  volume = {10},
  number = {1},
  pages = {5119},
  month = {nov},
  day = {11},
  issn = {2041-1723},
  doi = {10.1038/s41467-019-12963-3},
  url = {https://doi.org/10.1038/s41467-019-12963-3}
}

@article{Shinomiya2022,
  author = {Shinomiya, Hiroto and Sugimoto, Hiroshi and Hinamoto, Tatsuki and Lee, Yan Joe and Brongersma, Mark L. and Fujii, Minoru},
  title = {Enhanced Light Emission from Monolayer {MoS$_2$} by Doubly Resonant Spherical Si Nanoantennas},
  journal = {ACS Photonics},
  year = {2022},
  volume = {9},
  number = {5},
  pages = {1741--1747},
  month = {may},
  day = {18},
  publisher = {American Chemical Society},
  doi = {10.1021/acsphotonics.2c00142},
  url = {https://doi.org/10.1021/acsphotonics.2c00142}
}

@article{Katrisioti2025,
    author = {Katrisioti, Danae and Wiecha, Peter R. and Cuche, Aurélien and Psilodimitrakopoulos, Sotiris and Larrieu, Guilhem and Müller, Jonas and Larrey, Vincent and Urbaszek, Bernhard and Marie, Xavier and Stratakis, Emmanuel and Kioseoglou, George and Paillard, Vincent and Poumirol, Jean-Marie and Paradisanos, Ioannis},
    title = {Silicon nanoantennas for tailoring the optical properties of {MoS$_2$} monolayers},
    journal = {Appl. Phys. Lett.},
    volume = {127},
    number = {18},
    pages = {181101},
    year = {2025},
    month = {11},
    issn = {0003-6951},
    doi = {10.1063/5.0284138},
    url = {https://doi.org/10.1063/5.0284138}
}

@article{Poumirol2020,
  author = {Poumirol, Jean-Marie and Paradisanos, Ioannis and Shree, Shivangi and Agez, Gonzague and Marie, Xavier and Robert, Cédric and Mallet, Nicolas and Wiecha, Peter R. and Larrieu, Guilhem and Larrey, Vincent and Fournel, Frank and Watanabe, Kenji and Taniguchi, Takashi and Cuche, Aurélien and Paillard, Vincent and Urbaszek, Bernhard},
  title = {Unveiling the Optical Emission Channels of Monolayer Semiconductors Coupled to Silicon Nanoantennas},
  journal = {ACS Photonics},
  year = {2020},
  volume = {7},
  number = {11},
  pages = {3106--3115},
  month = {nov},
  day = {18},
  publisher = {American Chemical Society},
  doi = {10.1021/acsphotonics.0c01175},
  url = {https://doi.org/10.1021/acsphotonics.0c01175}
}

\begin{acknowledgments}
A.S. and C.X. fabricated the void resonators. G.G. and A.S. fabricated the heterostructures. PL measurements were performed by I.D.B., G.G., and A.S., while angle-resolved measurements were carried out by A.S. AFM measurements were conducted by A.S. COMSOL simulations were performed by A.S. and S.R. SEM measurements were performed by X.Z.-P.. Data processing, analysis, and plotting were done by A.S.. The original draft was written by A.S. and S.R. and revised by all co-authors. The project was supervised by S.R. with co-supervision by T.J.B. and A.H.
\end{acknowledgments}

\section*{Funding}
A.S. and S.R. acknowledge support by the Independent Research Fund Denmark (1032-00496B). S.R. and X.Z.-P. acknowledge support by Villum Fonden (VIL50376) and Novo Nordisk Foundation (NNF24OC0096142). G.G., I.D.B, A.H and T.J.B. acknowledges support by Novo Nordisk Foundation Challenge Program "BioMag" (Grant No. NNF21OC0066526).

\section*{Ethics declarations}
The authors declare no competing interests.
\end{document}